\documentclass[9pt,twocolumn,twoside]{pnas-new} 

\templatetype{pnasresearcharticle}

\title{Towards Bright Gamma-Ray Flash Generation From Tailored Target Irradiated by Multi-Petawatt Laser}

\author[a,1]{Prokopis Hadjisolomou}
\author[a]{Tae Moon Jeong}
\author[a,b]{Sergei V. Bulanov}

\affil[a]{ELI Beamlines Centre, Institute of Physics, Czech Academy of Sciences, Za Radnicí 835, 25241 Dolní Břežany, Czech Republic}
\affil[b]{National Institutes for Quantum and Radiological Science and Technology (QST), Kansai Photon Science Institute, 8-1-7 Umemidai, Kizugawa, Kyoto 619-0215, Japan}

\leadauthor{Hadjisolomou} 

\significancestatement{Gamma-ray flashes are phenomena of diachronic interest that occur not only in the microcosm but also in cosmic environments. In the laboratory, extreme gamma-flashes may be generated during laser-matter interaction. In finding optimal conditions of $\gamma$-photon generation, it is crucial to elucidate the role of laser pulse structure comprising of ultrashort main pulse, preceded by a relatively long background field that alters the target profile. Background field removal not only adds complexity to high power lasers but also comes at cost of reduced laser energy. In the present work, we find an approach for substantial enhancement of the laser energy conversion to $\gamma$-photon energy.}

\authorcontributions{Author contributions: P.H., T.M.J. and S.V.B. conceived of the work and on interpretation of the results; P.H. wrote the paper, performed PIC simulations and analyzed the data.}

\authordeclaration{The authors declare no competing interest.}

\correspondingauthor{\textsuperscript{1}To whom correspondence should be addressed. E-mail: prokopis.hadjisolomou@eli-beams.eu}


\begin{abstract}
One of the remarkable phenomena in the laser-matter interaction is the extremely efficient energy transfer to $\gamma$-photons, that appears as a collimated $\gamma$-ray beam. For interactions of realistic laser pulses with matter, existence of a background field plays a crucial role, since it hits the target prior to the main pulse arrival, leading to a cloud of preplasma and drilling a narrow channel inside the target. These effects significantly alter the process of $\gamma$-photon generation. Here, we study this process by importing the outcome of magnetohydrodynamic simulations of the target interaction into particle-in-cell simulations for describing the $\gamma$-photon generation. It is seen that the background field effect plays an important positive role, enhancing the efficiency of laser pulse coupling with the target, and generating high energy electron-positron pairs. It is expected that such a $\gamma$-photon source will be actively used in various applications in nuclear photonics, material science and astrophysical processes modeling.
\end{abstract}

\dates{This manuscript was compiled on \today}
\doi{\url{www.pnas.org/cgi/doi/10.1073/pnas.XXXXXXXXXX}}

\begin{document}

\maketitle
\thispagestyle{firststyle}
\ifthenelse{\boolean{shortarticle}}{\ifthenelse{\boolean{singlecolumn}}{\abscontentformatted}{\abscontent}}{}


\dropcap{S}ince the invention of the Chirped Pulse Amplification technique \cite{1985_StricklandD}, development of multi-petawatt (multi-PW) laser systems is envisioned worldwide. The first laser to reach the $10 \kern0.2em \mathrm{PW}$ peak power level has been recently reported by ELI-NP \cite{2020_TanakaKA}, Romania, with a pulse duration of ${\sim} \kern0.1em 25 \kern0.2em \mathrm{fs}$. Another ${\sim} \kern0.1em 10 \kern0.2em \mathrm{PW}$ laser is soon expected in ELI-Beamlines \cite{2019_DansonC}, Czech Republic, with approximately five times higher energy. In addition, ELI-ALPS, Hungary, aims at constructing an ultrashort ${\sim} \kern0.1em 17 \kern0.2em \mathrm{fs}$ laser of ${\sim} \kern0.1em 2 \kern0.2em \mathrm{PW}$ \cite{2019_Osvay}. Moreover, a laser combining pulses shorter than $20 \kern0.2em \mathrm{fs}$ in the $10 \kern0.2em \mathrm{PW}$ level is developed in Apollon facility \cite{2016_PapadopoulosDN}, France. By focusing a multi-PW laser down to a micrometer-wide spot \cite{2018_KiriyamaH, 2021_YoonJ}, intensities exceeding $10^{27} \kern0.2em \mathrm{W \kern0.1em m^{-2}}$ can be achieved, where this threshold has been recently surpassed by the ${\sim} \kern0.1em 4 \kern0.2em \mathrm{PW}$ CoReLS laser \cite{2021_YoonJ}, South Korea.

\par A typical high power laser consists of an ultrashort main pulse, preceded by a lower amplitude background field extending in the nanosecond scale. The laser contrast is defined as the ratio of the main pulse amplitude to the background field amplitude. Usually, in high power laser systems the contrast is increased through complex and/or expensive additions, such as Optical Parametric Chirped Pulse Amplification \cite{DUBIETIS1992437} and plasma mirrors \cite{2006_MourouG, 2007_LevyA}.

\par The importance of a finite contrast for laser-matter interactions is highlighted both experimentally \cite{2002_GiuliettiD, 2003_MatsukadoK, 2008_YogoA, 2012_OguraK} and theoretically \cite{2003_MatsukadoK, 2004_UtsumiT, 2008_YogoA, 2014_EsirkepovTZ, Hadjisolomou2020}. By assuming an initially steep density gradient (flat-foil, or simply foil) target, in the aforementioned literature it is agreed that the background field modifies the initial density profile at an extent proportional to its amplitude and duration. A relatively thick (micrometer range) foil is curved in the vicinity of the laser focal spot, where a gradually increasing density profile appears in the target front region. On the other hand, if the target is thin enough then the background field drills the target resulting in no interaction when the main pulse arrives. As a result, the preplasma strongly affects the energy spectra and directionality of particles emitted due to the laser-target interaction. However, since computational studies of the interaction typically involve particle-in-cell (PIC) simulations which cannot be applied to model the nanosecond long duration required by the background field, usually a foil target is assumed, acknowledging only the effect of the main pulse on target.

\par A plethora of exotic target geometries has been already considered for laser-matter interaction experimnets. Among numerous examples, we mention proton-rich micro-dots \cite{2006_SchwoererH}, cylindircal micro-lenses \cite{2006_ToncianT, 2008_KarS}, hollow micro-spheres \cite{2011_BurzaM}, micro-coils \cite{2016_KarS} and wavelength-scale holed targets \cite{2016_PSikalJ, 2020_HadjisolomouP}. All of the aforementioned target designs require explicit microfabrication techniques, while they add further complexity to a laser-target experiment since they require additional efforts on positioning and alignment of the target. However, it was noticed that the use of tailored targets is favorable for the laser-target interaction and their use is widely employed.

\par Apart of target tailoring methods outside the interaction chamber, it is also possible to manipulate the electron density profile by means of a secondary, long pulse duration, lower power laser \cite{1986_GitomerSJ}. The secondary laser provides the required background field, where temporal control of the laser \cite{2022_DorrerC} can provide the required electron density distributions. This scheme is known as `laser heater' and has been successfully implemented on improving the properties of laser driven particle beams \cite{1996_BorghesiM, 2021_GizziLA}. Notably, manipulating the target density with a laser heater resulted in record energy values of particles accelerated by optical means \cite{2019_GonsalvesAJ}. 

\par At the ultrahigh intensity limit, the theory foresees that the multi-PW lasers interacting with matter will provide not only energetic charged particles (electrons ($\mathrm{e^-}$), protons ($\mathrm{p^+}$), heavy ions ($\mathrm{i^+}$)) but also a plethora of high energy $\gamma$-photons and electron-positron ($\mathrm{e^- \mbox{-} e^+}$) pairs. Once an electron (or positron) collides with the incident field it is scattered, resulting in alteration of its momentum and a $\gamma$-photon is emitted, in a process known as multiphoton Compton scattering. In the present work, we focus on $\gamma$-photons produced via the multiphoton Compton scattering dominating at ultrahigh intensities \cite{2012_NakamuraT, 2013_RidgersCP, 2018_LezhninKV, 2021_YounisAH}, produced within a time approximately equal to to the laser pulse duration. Notably, at lower intensities ($< 10^{27} \kern0.2em \mathrm{W \kern0.1em m^{-2}}$) and significantly thick targets $\gamma$-photons can be also produced via Bremsstrahlung emission \cite{1959_KochHW}, where the present target geometry limits Bremsstrahlung contribution. The $\gamma$-photon production then makes possible the production of $\mathrm{e^- \mbox{-} e^+}$ pairs via the multiphoton Breit-Wheeler process \cite{2009_EhlotzkyF}, where a high energy $\gamma$-photon interacts with multiple laser photons. Maximizing the laser to $\gamma$-photon energy conversion efficiency, $\kappa_\gamma$, is required in photonuclear physics \cite{2000_LedinghamKW, 2004_NedorezovVG}, for study of high energy density physics in materials science \cite{2013_EliassonB} and studies of astrophysical processes \cite{1992_ReesMJ, 2015_BulanovSV, 2018_PhilippovAA, 2021_AharonianF}.

\par The recently available multi-PW lasers in combination with the broad application range of the $\gamma$-photons, draw the interest of several research groups, suggesting various methods to maximize $\kappa_\gamma$. An early suggestion is based on the reflected part of the laser pulse incident on an overdense target \cite{2002_ZhidkovA, 2005_KogaJ, 2018_GuYJ}. However, loss of significant laser energy towards the reflection region suggested the use of two counter-propagating pulses \cite{2008_BellAR, 2009_KirkJG, 2015_LuoW, 2016_GrismayerT}, later extending the scheme to multiple colliding pulses \cite{2016_VranicM, 2017_GongZ}. Apart from the all-optical approach, other groups suggested microfabricating sophisticated target schemes \cite{2019_JiLL, 2021_ZhangLQ} or even combining a laser with either optically \cite{2014_SarriG} or externally accelerated electrons \cite{2019_MagnussonJ}. Moreover, it was shown that by adding a chosen preplasma on the target front, $\kappa_\gamma$ can also be increased \cite{2018_LezhninKV, 2020_WangXB}. 


\section*{Hybrid Magnetohydrodynamic and Particle-in-Cell Simulations}

\par In reference \cite{2022_TsygvintsevIP} it has been demonstrated through magnetohydrodynamic (MHD) simulations that a unique microfabrication of the foil target is achieved by a background field corresponding to multi-PW lasers. The radial symmetry of the results allows their expansion on a $({\mathbf{\hat{x}}}, {\mathbf{\hat{y}}}, {\mathbf{\hat{z}}})$ grid, where the laser propagation axis is set along ${\mathbf{\hat{x}}}$. Therefore one can obtain the three-dimensional (3D) electron number density distribution. The results of the MHD simulations are then used as initial conditions in PIC simulations, in a hybrid modeling scheme that reveals the physical processes governing the complete interaction of the main laser pulse with the newly formed density distribution.

\par The background field sets a $60 \kern0.2em \mathrm{ps}$ duration region with intensity starting at $10^{19} \kern0.2em \mathrm{W \kern0.1em m^{-2}}$, dropping exponentially to $10^{14} \kern0.2em \mathrm{W \kern0.1em m^{-2}}$, and then continues with a constant intensity for $940 \kern0.2em \mathrm{ps}$. The laser intensity spatial profile is represented by a Gaussian of ${\sim} \kern0.1em 2.2 \kern0.2em \mathrm{\mu m}$ full-width-at-half-maximum. The aforementioned background field intensity is comparable with what was considered as cutting-edge laser main pulse intensity a few decades ago \cite{1986_GitomerSJ}. From reference \cite{2022_TsygvintsevIP} we choose a set of elements with approximately equally increasing electron number density, $n_e$, namely lithium (Li, $n_e \approx 1.39 \kern0.1em {\times} \kern0.1em 10^{29} \kern0.2em \mathrm{m^{-3}}$), sodium (Na, $n_e \approx 2.79 \kern0.1em {\times} \kern0.1em 10^{29} \kern0.2em \mathrm{m^{-3}}$), beryllium (Be, $n_e \approx 4.95 \kern0.1em {\times} \kern0.1em 10^{29} \kern0.2em \mathrm{m^{-3}}$), carbon (C, $n_e \approx 6.02 \kern0.1em {\times} \kern0.1em 10^{29} \kern0.2em \mathrm{m^{-3}}$), and aluminum (Al, $n_e \approx 7.83 \kern0.1em {\times} \kern0.1em 10^{29} \kern0.2em \mathrm{m^{-3}}$).

\begin{figure}[hb]
\centering
\includegraphics[width=0.8\linewidth]{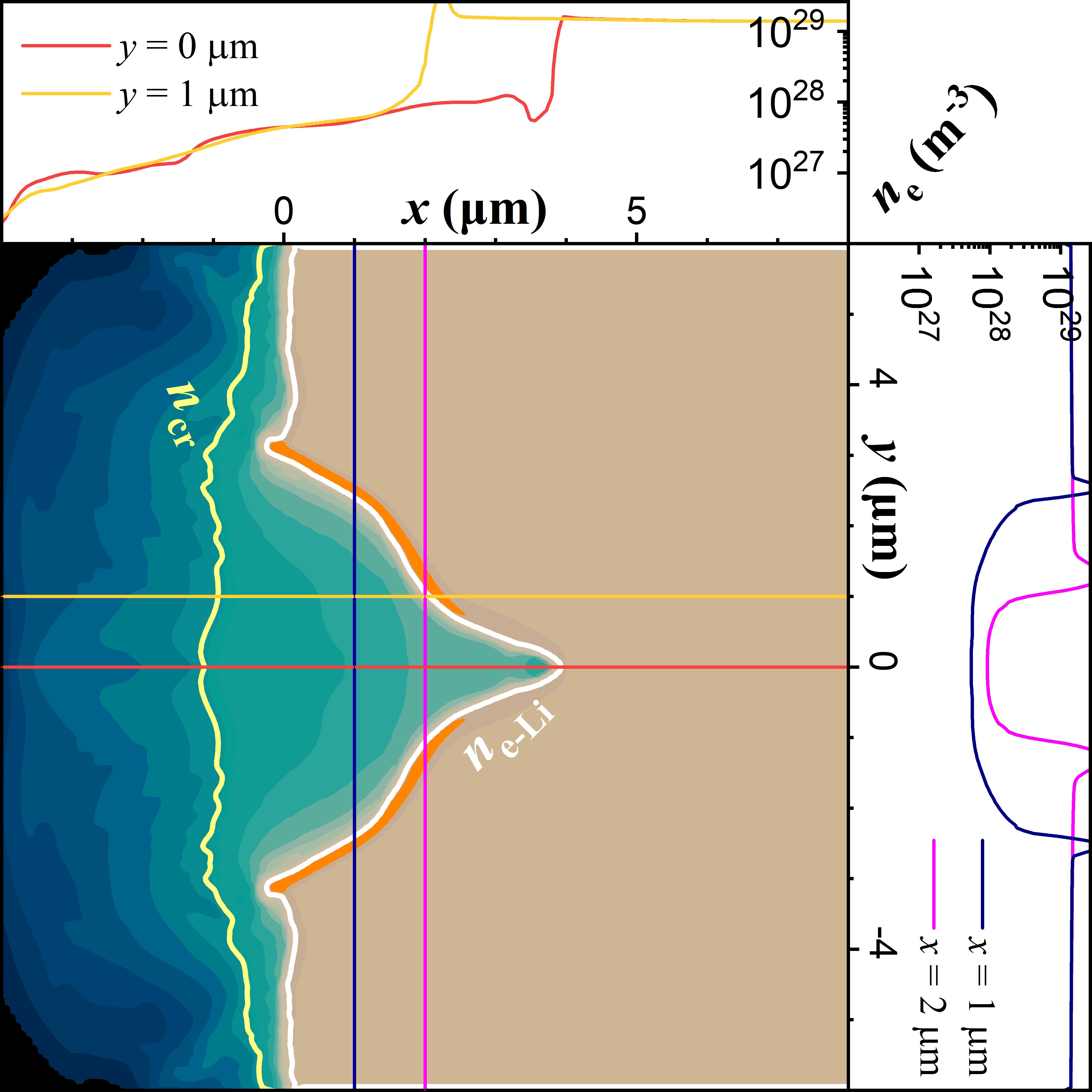}
\caption{Electron number density as given by MHD simulations (data taken from reference \cite{2022_TsygvintsevIP}), following background laser field irradiation of a lithium foil. The yellow contour line is at the critical density and the white contour line is at the lithium solid electron density. The orange saturated contour is overcritical for laser intensities above $10^{27} \kern0.2em \mathrm{W \kern0.1em m^{-2}}$.}
\label{fig:fig1}
\end{figure}

\par The effect of the background field on lithium is shown in Fig. \ref{fig:fig1}, where lithium having the lowest density among all solids at room temperature it is affected the strongest. In agreement with previous works \cite{2014_EsirkepovTZ, 2018_LezhninKV}, a preplasma distribution is generated in the target front region, although with a large exponential coefficient for the electron number density gradient of ${\sim} \kern0.1em 3.1 \kern0.1em {\times} \kern0.1em 10^5 \kern0.2em \mathrm{m^{-1}}$, shown by the red lineout in Fig. \ref{fig:fig1}, along the laser propagation axis. Notably, the preplasma exceeding the critical density, $n_{cr}$, extends for only ${\sim} \kern0.1em 1 \kern0.2em \mathrm{\mu m}$ in the target front region.

\par The sizable target ($12 \kern0.2em \mathrm{\mu m}$ diameter and $10 \kern0.2em \mathrm{\mu m}$ thickness) ensures neither target curvature on the target rear, nor complete drilling and destruction of the target \cite{2014_EsirkepovTZ, Hadjisolomou2020}. In addition to the preplasma formation, a conical-like cavity is generated within the target volume, with walls several times denser than the background. For lithium, the cavity has a depth of ${\sim} \kern0.1em 4 \kern0.2em \mathrm{\mu m}$ and a base diameter of ${\sim} \kern0.1em 6 \kern0.2em \mathrm{\mu m}$. The Gaussian background field profile allows stronger target deformation in the center, where the cavity profile exhibits two off-center symmetric vertexes (where the derivative of the curvature is zero). This unique tailored target is achievable with finite contrast lasers currently available, revealing a preplasma regime which to the best of our knowledge has never been considered in PIC simulations.

\par The PIC code used is the relativistic quantum electrodynamic EPOCH code \cite{2015_ArberTD},  compiled with the Higuera-Cary \cite{Higuera2017} (to more accurately resolve electron trajectories), Bremsstrahlung and Photons \cite{2014_RidgersCP} prepocessor directives enabled. The simulations are performed in the 3D version of the code, where the laser pulse and target characteristics match those of the MHD simulations, with main pulse focused intensity ranging from ${\sim} 2.8 \kern0.1em {\times} \kern0.1em 10^{26} \kern0.2em \mathrm{W \kern0.1em m^{-2}}$ to ${\sim} 2.8 \kern0.1em {\times} \kern0.1em 10^{27} \kern0.2em \mathrm{W \kern0.1em m^{-2}}$. The corresponding energy is set from $20 \kern0.2em \mathrm{J}$ to $200 \kern0.2em \mathrm{J}$, while a pulse duration of ${\sim} \kern0.1em 17 \kern0.2em \mathrm{fs}$ \cite{2019_Osvay, 2021_YoonJ} corresponds to lasers of ${\sim} \kern0.1em 1 \kern0.2em \mathrm{PW}$ to ${\sim} \kern0.1em 10 \kern0.2em \mathrm{PW}$ power. The central laser wavelength is set to $815 \kern0.2em \mathrm{nm}$, typical for titanium-sapphire lasers. The laser focal spot is set at the center of the front surface of a foil target.

\par The MHD density array, after interpolation to a $10 \kern0.1em {\times} \kern0.1em 40 \kern0.1em {\times} \kern0.1em 40 \kern0.2em \mathrm{nm}$ cell size grid, is centered and expanded in a cubic PIC volume, extending from $-15.36 \kern0.2em \mathrm{\mu m}$ to $15.36 \kern0.2em \mathrm{\mu m}$ in all three directions. The cell dimensions are chosen small enough to accurately resolve the relativistic skin depth, while 8 macro-electrons and 8 macro-ions are assigned to each cell. The simulation ran for $110 \kern0.2em \mathrm{fs}$, enough for $\kappa_\gamma$ to saturate and at the same time neither field energy nor energetic electrons escape the simulation box from the open boundaries. The laser is focusing at a simulation time of ${\sim} \kern0.1em 65 \kern0.2em \mathrm{fs}$, which is set as $0 \kern0.2em \mathrm{fs}$ for the laser-target interaction.


\section*{Cavity Propagation and Intensity Enhancement}

\begin{figure}[h]
\centering
\includegraphics[width=0.77\linewidth]{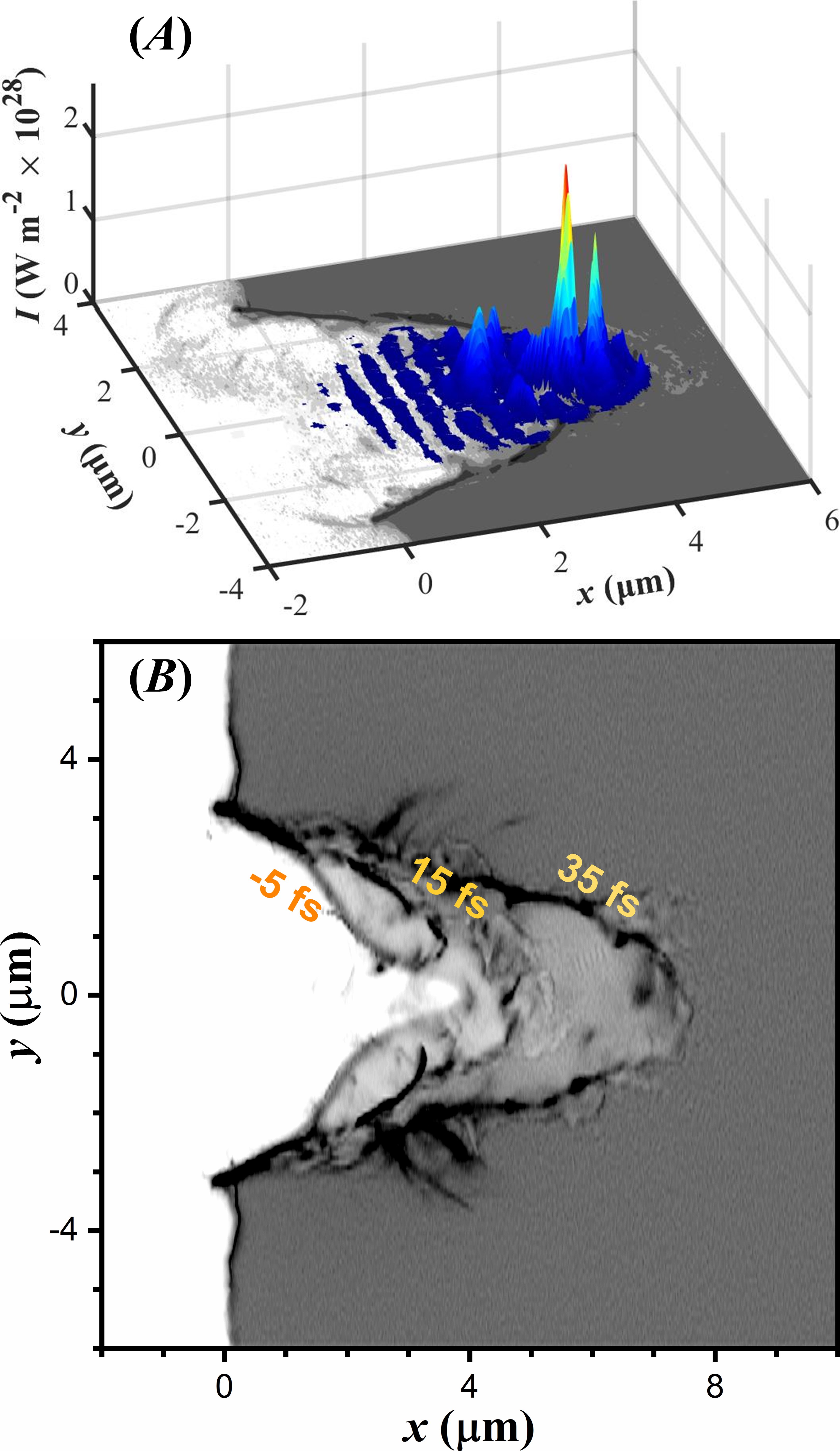}
\caption{(\textit{A}) Laser intensity (color surface plot) overlayed to the lithium electron number density (grayscale image) at ${\sim} \kern0.1em  15 \kern0.2em \mathrm{fs}$, when an intensity of ${\sim} \kern0.1em 2.6 \kern0.1em {\times} \kern0.1em10^{28} \kern0.2em \mathrm{W \kern0.1em m^{-2}}$ is reached. (\textit{B}) Overlay of three successive electron number density distributions, with a time step of $20 \kern0.2em \mathrm{fs}$. The first layer is at ${\sim} \kern0.1em -5 \kern0.2em \mathrm{fs}$, when the main pulse is within the cavity. The overlay of the three layers reveals the temporal dynamics of the cavity formation within the target.}
\label{fig:fig2}
\end{figure}

\par In the present work, the default target referral is a tailored lithium target and the default laser power is ${\sim} \kern0.1em 10 \kern0.2em \mathrm{PW}$, except where it is stated otherwise. As the background field deposits most of its energy on target, a conical-like cavity is generated prior to the arrival of the main laser pulse. This scheme resembles laser-target interaction geometries on which a cone is purposely fabricated at the target, aiming at novel fast-ignition schemes \cite{2002_KodamaR, 2021_KambojO}, increasing laser induced $\gamma$-photon production \cite{2022_ChintalwadS}, enhancing laser field intensity \cite{2020_BudrigaO}, and efficient proton acceleration \cite{2012_BadziakJ, 2013_BusoldS}.

\begin{figure}[h]
\centering
\includegraphics[width=0.75\linewidth]{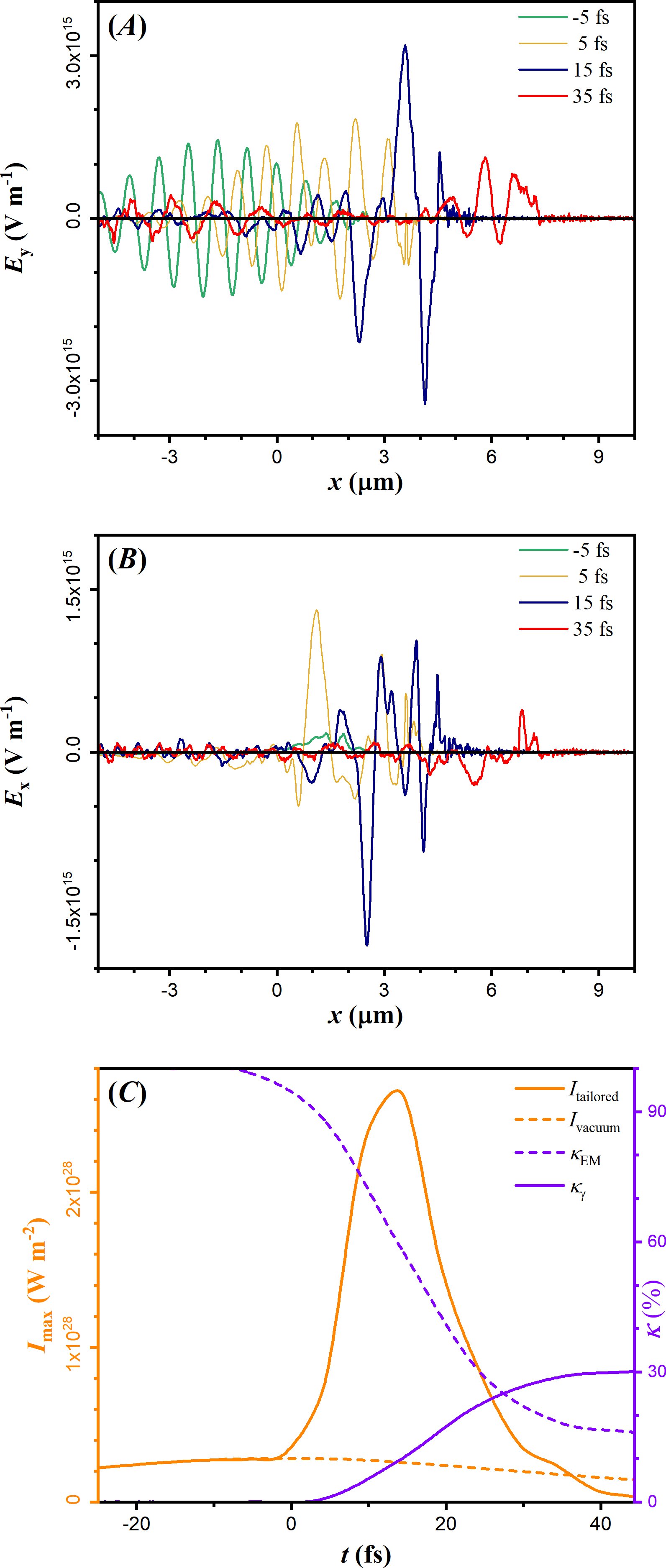}
\caption{(\textit{A}) Line-out of $E_y$, and (\textit{B}) line-out of $E_x$ along the laser propagation axis at various times, as noted on the legend. Note that the laser is linearly polarized, along $\mathbf{\hat{y}}$. (\textit{C}) The left axis shows the maximum field intensity as a function of simulation time; the black dashed baseline denotes the field intensity expected in the focal spot, without the target. The right axis shows with the dashed purple line the amount of laser energy transferred to all target particles, while the solid purple line shows the percentage of the remaining laser field energy, $\kappa_\gamma$, as a function of time.}
\label{fig:fig3}
\end{figure}

\par Since in our case the relativistic critical density is above lithium electron number density, one might expect the cavity formation to have no effect on focusing the laser field, due to dominance of relativistic self-focusing \cite{2001_FeitMD} of the laser field in the underdense solid. However, as seen by the spikes in the density lineout profiles in Fig. \ref{fig:fig1}, a thin, overdense layer is formed on the cavity walls. As a result, the laser field can penetrate only within the skin-depth of the walls, and reflected towards the cavity depth.

\par The reflected fields then interfere, rapidly increasing the field intensity. Note that the laser focal spot is at the base of the cone, meaning that in vacuum the laser would be defocusing if the cavity did not exist. In addition, the cavity volume is filled with a low electron density, that also aids focusing due to weak relativistic self-focusing; although to a lesser extent than the cavity focusing, as its electron density increases exponentially.

\par Once the laser is intensified in the cavity, its intensity surpasses that expected in the focal spot. As a result, the thin overdense electron distribution becomes undercritical, and the laser breaks into the target volume \cite{2005_ManglesSPD}, where the lithium density is not modified by the background field. In addition, field reflection by the cavity walls results in caustics, instantaneously increasing the field intensity. These effects are seen in Fig. \ref{fig:fig2}(\textit{A}), where the laser intensity is shown when reaching its maximum value of ${\sim} \kern0.1em 2.6 \kern0.1em {\times} \kern0.1em10^{28} \kern0.2em \mathrm{W \kern0.1em m^{-2}}$, an order of magnitude higher than the intensity expected at the focal spot. The electron number density is overplotted on the figure, visualizing the spatial location of the laser field with respect to the cavity. The highest intensity is recorded at ${\sim} \kern0.1em 15 \kern0.2em \mathrm{fs}$ after the laser field reaches the focal spot.

\par Figs. \ref{fig:fig3}(\textit{A})  and \ref{fig:fig3}(\textit{B}) show line-outs (along laser propagation axis) of the transverse and longitudinal components of the electric field ($\mathbf{E_y}$ and $\mathbf{E_x}$ respectively) at successive times. When the laser field arrives at the cavity (at ${\sim} \kern0.1em -5 \kern0.2em \mathrm{fs}$), $\mathbf{E_x}$ is practically absent. However, reflection of the laser field on the cavity walls allows not only interference of $\mathbf{E_y}$ but also generation of $\mathbf{E_x}$, where the magnitude of the two field components is comparable since $E_y \approx 2 E_x$ for times after ${\sim} \kern0.1em 0 \kern0.2em \mathrm{fs}$. Existence of $\mathbf{E_x}$ ceases only when both field components dissipate, near the end of the simulation.

\par Penetration of the laser field in the target at successive times is shown in Fig. \ref{fig:fig2}(\textit{B}), by overlaying three electron density contours. The color magnitude serves purely for visualization purposes, since appropriate contrast is applied to the figure to reveal the cavity propagation. Although the laser can penetrate the relativistically undercritical target, its propagation is further assisted by $\mathbf{E_x}$, which further drills the target along its propagation axis. 

\par  As a result, the laser pulse can maintain high intensities until the end of the simulation, although dissipating due to energy transfer to electrons and $\gamma$-photons. This can be seen in Fig. \ref{fig:fig3}(\textit{C}), where the left axis plots the laser peak intensity as a function of time. For comparison, the orange dashed line is the intensity in the focal spot in vacuum. At ${\sim} \kern0.1em -5 \kern0.2em \mathrm{fs}$ the laser field is in a region of extremely low electron density, also in the vicinity of the focal spot region. Therefore, the peak intensity coincides with the expected peak intensity in vacuum. Beyond that time, up to ${\sim} \kern0.1em 15 \kern0.2em \mathrm{fs}$, the intensity is increasing. At that time interval most of the field energy is transferred to the rest of particles ($\gamma$-photons, electrons, positrons, ions), shown by the dashed purple line in Fig. \ref{fig:fig3}(\textit{C}). 

\par Laser energy conversion to particle energy continues up to ${\sim} \kern0.1em 35 \kern0.2em \mathrm{fs}$, where the peak intensity drops since no significant laser energy remains within the propagating cavity. The laser field energy saturates at ${\sim} \kern0.1em 15 \kern0.2em \%$ of its initial value, due to laser back-reflection. The $\kappa_\gamma$, shown by the solid purple line, exhibits a sigmoidal behavior with changing curvature at the time the laser intensity is maximized, saturating at ${\sim} \kern0.1em 30 \kern0.2em \%$. Remarkably, conversion of approximately one third of the laser energy to $\gamma$-photons with the currently available technology, is of interest for worldwide laser facilities.


\section*{Gamma-Ray Flash Scaling with Power and Target Material}

\par
\begin{figure}[bh]
\centering
\includegraphics[width=0.87\linewidth]{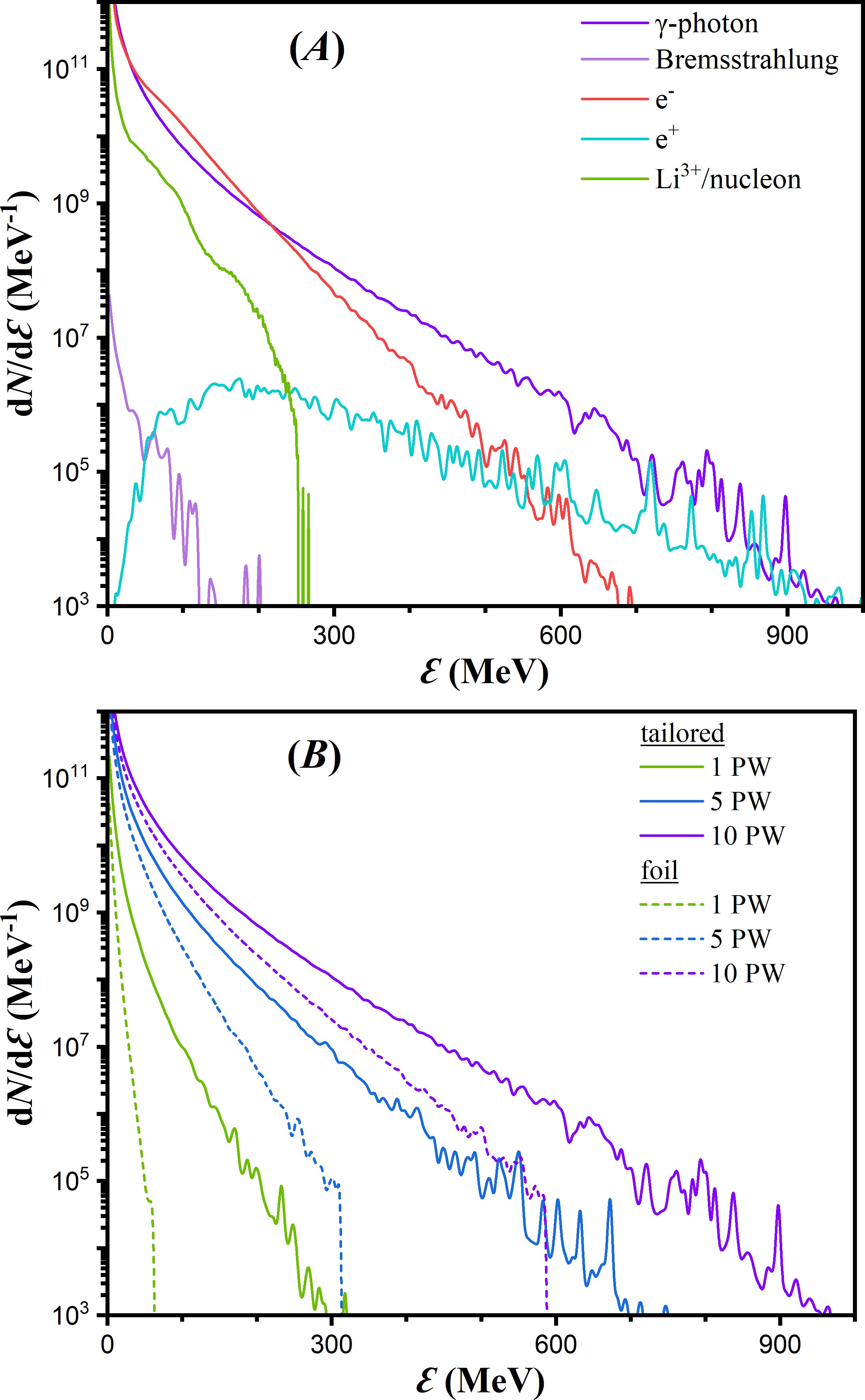}
\caption{(\textit{A}) Energy spectra after a ${\sim} \kern0.1em 10 \kern0.2em \mathrm{PW}$ laser interacts with a tailored lithium target. (\textit{B}) Energy spectra of $\gamma$-photons after interaction of the lithium target with main laser pulses of varying power, as noted on the legend. The solid lines correspond to tailored targets, and the dashed lines to foil targets.}
\label{fig:fig4}
\end{figure}

 The aforementioned  laser-target interaction leads to a plethora of energetic electrons in the GeV-scale, which in turn result in ion acceleration with energies per nucleon of ${\sim} \kern0.1em 300 \kern0.2em \mathrm{MeV}$ as seen in Fig. \ref{fig:fig4}(\textit{A}). We should clarify that optimizing ion acceleration is not among the goals of the present work, hence the thick target chosen. By considering that for optimal laser coupling to electrons \cite{1998_VshivkovVA} the condition $a_0=\pi (n_e/n_{cr}) (l/\lambda)$ must be satisfied (where $a_0$ is the dimensionless amplitude and $l$ is the penetration depth per wavelength), it can be calculated that the intensity ranges shown in Fig. \ref{fig:fig3}(\textit{C}) correspond to $1 \kern0.2em \mathrm{\mu m} < l< 3 \kern0.2em \mathrm{\mu m}$. Since the temporal pulse length is ${\sim} 6 \kern0.1em \lambda$, it is estimated that a ${\sim} \kern0.1em 10 \kern0.2em \mathrm{\mu m}$ thick target is required for the laser pulse to efficiently transfer its energy to the target.

\par In the present work we consider $\gamma$-photons produced by either Bremsstrahlung radiation or multiphoton Compton scattering. However, the micrometer-thick, low atomic number target is unsuitable for Bremsstrahlung emission, although the emission continues for a significantly longer time than the pulse duration. In addition, the stochastic dynamics of electrons in the solid target results in a non-directional Bremsstrahlung emission, making $\gamma$-photons from Compton scattering dominate the $\gamma$-photon signal during ultraintense laser-matter experiments. The Compton $\gamma$-photon spectrum is comparable to the electron spectrum, as seen in Fig. \ref{fig:fig4}(\textit{A}), with a temperature of ${\sim} \kern0.1em 55 \kern0.2em \mathrm{MeV}$ for $\gamma$-photons of $> 200 \kern0.2em \mathrm{MeV}$ energy. For comparison, the Bremsstrahlung spectrum at the end of the simulation is also shown, which is several orders of magnitude lower than the Compton spectrum. However, its contribution can only increase the cumulative $\gamma$-photon energy, with a  $\kappa_\gamma$ of ${\sim} \kern0.1em 30 \kern0.2em \%$ only from Compton scattering contribution. Hence, the default $\gamma$-photon referral is Compton scattering.

\par At high intensity laser-matter interactions, generation of $\mathrm{e^- \mbox{-} e^+}$ pairs is possible through the multiphoton Breit-Wheeler process. Although $\mathrm{e^- \mbox{-} e^+}$ pair generation is not expected to be observed for a ${\sim} \kern0.1em 10 \kern0.2em \mathrm{PW}$ laser (corresponding to intensities of ${\sim} 2.8 \kern0.1em 10^{27} \kern0.2em \mathrm{W \kern0.1em m^{-2}}$), we demonstrate that the one order of magnitude intensity enhancement due to the tailored target makes multiphoton Breit-Wheeler $\mathrm{e^- \mbox{-} e^+}$ pair observation feasible. The positron spectrum is shown in Fig. \ref{fig:fig4}(\textit{A}), spanning up to the maximum $\gamma$-photon energy with a Maxwell-Juttner distribution of ${\sim} \kern0.1em 80 \kern0.2em \mathrm{MeV}$ temperature, where ${\sim} \kern0.1em 4 \kern0.1em {\times} \kern0.1em 10^8$ pairs are recorded.

\par
\begin{figure}[h]
\centering
\includegraphics[width=0.9\linewidth]{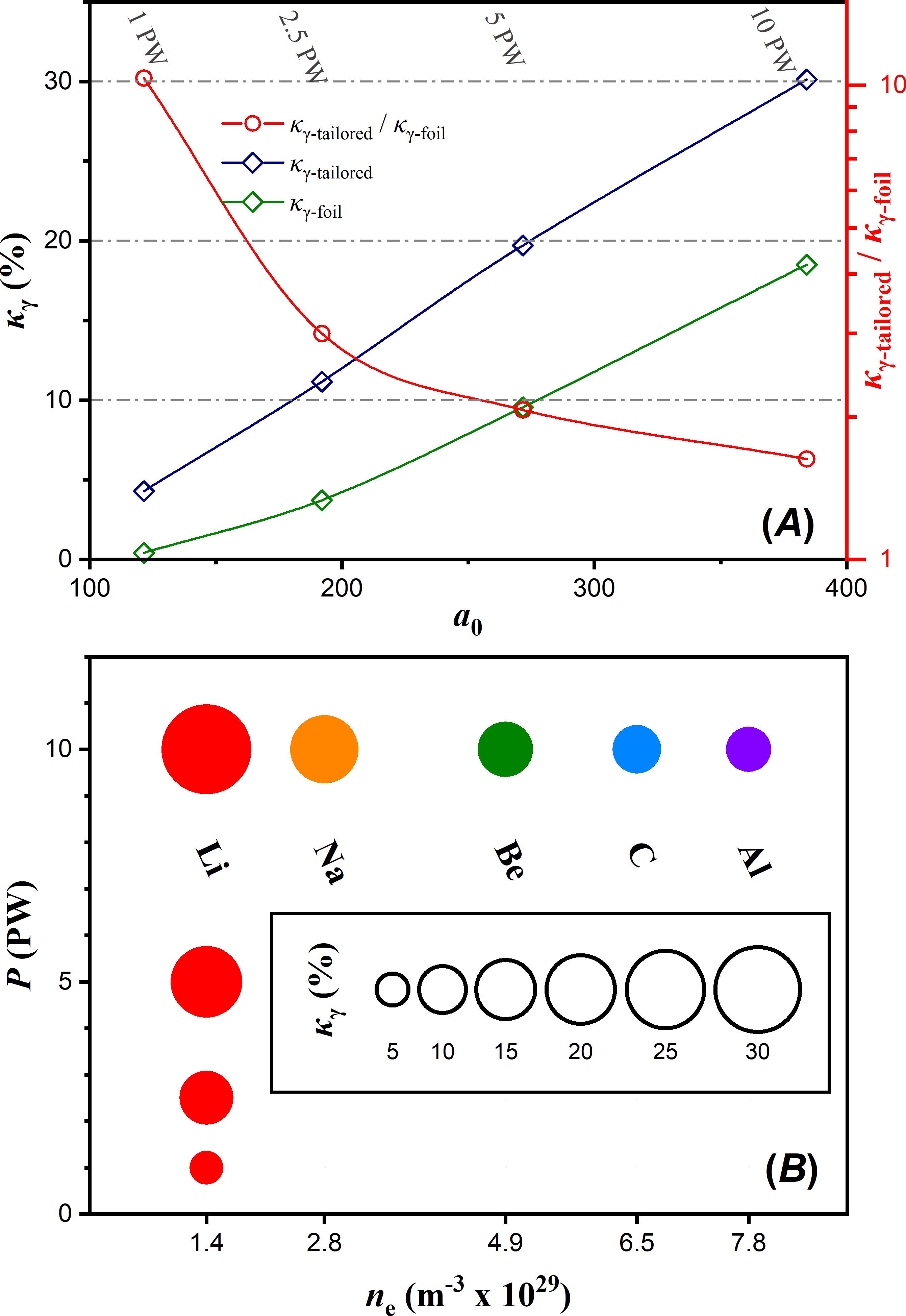}
\caption{(\textit{A}) The left axis shows $\kappa_\gamma$ as a function of $a_0$, for tailored (blue) and foil (green) lithium targets. The right axis shows the ratio of the two aforementioned cases. (\textit{B}) $\kappa_\gamma$ as a function of laser power, for various tailored target materials. The $\kappa_\gamma$ value is proportional to the circle area.}
\label{fig:fig5}
\end{figure}

 The preformed cavity importance on $\kappa_\gamma$ optimization and $\gamma$-photon maximum energy increase is seen in Fig. \ref{fig:fig4}(\textit{B}) by comparing the solid lines (tailored target) with the dashed lines (foil target) for various laser powers. In all cases, the cavity formation significantly amplifies the $\gamma$-photon spectrum, where the exponential behavior still persists but with decreasing temperature for decreased power. The quantitative results for $\kappa_\gamma$ as a function of $a_0$ are shown in Fig. \ref{fig:fig5}(\textit{A}), where the blue line corresponds to tailored targets and the green to foil targets. The figure reveals an almost linear dependency of $\kappa_\gamma$ to $a_0$  (or to the laser power) within the range of interest.

\par The right axis of Fig. \ref{fig:fig5}(\textit{A}) shows the ratio of $\kappa_\gamma$ from tailored to that of foil lithium targets as a function of $a_0$. Notably, for a ${\sim} \kern0.1em 10 \kern0.2em \mathrm{PW}$ laser the ratio is ${\sim} \kern0.1em 1.5$, but rising to ${\sim} \kern0.1em 10$ for a ${\sim} \kern0.1em 1 \kern0.2em \mathrm{PW}$ laser. This result is related to the relativistic transparency of lithium for the ${\sim} \kern0.1em 10 \kern0.2em \mathrm{PW}$ case, where even in the case of no cavity formation the laser pulse is still efficiently penetrating the target. On the other hand, the ${\sim} \kern0.1em 1 \kern0.2em \mathrm{PW}$ case relies on intensity enhancement within the cavity to reach the relativistic transparency threshold, whilst the foil target strongly reflects the laser pulse with little-to-no conversion to $\gamma$-photons. Therefore, even single-PW lasers can be used to efficiently create a $\gamma$-ray flash in the laboratory, when a proper background field is combined with a relatively low density target.

\par In connection to the high $\kappa_\gamma$ for lithium targets, the results for denser materials are shown in the columns of \ref{fig:fig5}(\textit{B}), where the rows correspond to varying laser power. Study of the dependency of $\kappa_\gamma$ on the target material is performed only for the ${\sim} \kern0.1em 10 \kern0.2em \mathrm{PW}$ case, since in all lower powers it follows a similar trend. For the materials considered, it is found that $\kappa_\gamma$ is inversely proportional to $n_e$. By increasing the target density, two effects result in reduced $\kappa_\gamma$. Firstly, the cavity formation is less prominent in a denser material, resulting in lower intensity amplification. Secondly, the target electron number density shifts away from the relativistic critical density and the laser field can no longer be efficiently coupled to the target electrons. Notably, materials commonly used in experiments have an electron density of a few times higher than that of lithium (e.g. approximately five times for aluminum). As a result, if aluminum is used in a ${\sim} \kern0.1em 10 \kern0.2em \mathrm{PW}$ $\gamma$-ray flash experiment, $\kappa_\gamma$ will be significantly suppressed. The main drawback of lithium is its high chemical reactivity in air. Fortunately, a thin (sub-micron) polymer coating does not significantly alter the laser interaction with the target since it is drilled by the background field before the main pulse arrives. In addition, the laser-target interaction takes place in vacuum.


\section*{Gamma-Ray Flash Parameters}

\begin{figure}[h]
\centering
\includegraphics[width=0.97\linewidth]{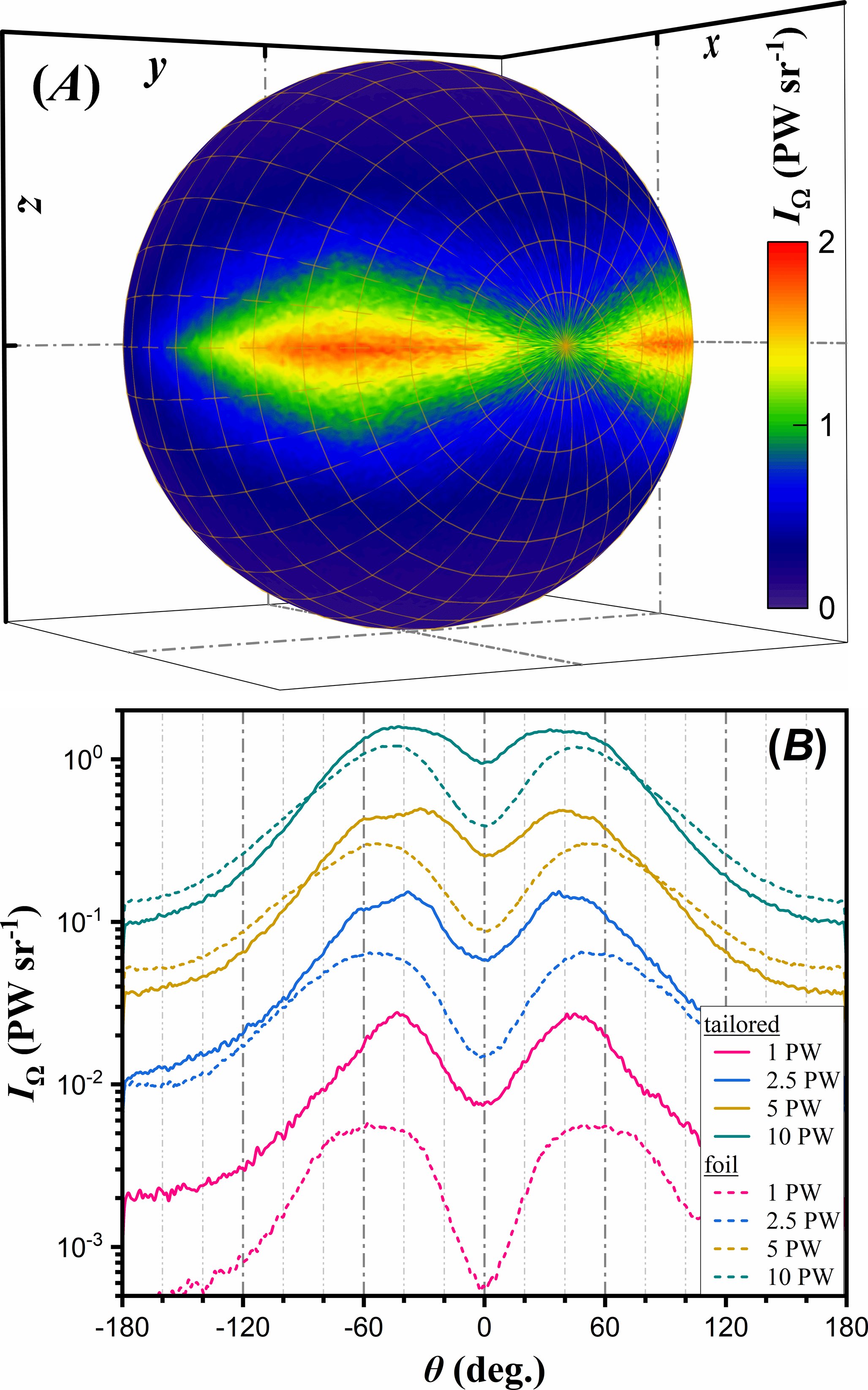}
\caption{(\textit{A}) Radiant intensity of $\gamma$-photons due to the interaction of a ${\sim} \kern0.1em 10 \kern0.2em \mathrm{PW}$ laser with a tailored lithium target. (\textit{B}) Line-out of the radiant intensity of $\gamma$-photons, taken along the polarization plane and within a full-angle divergence of $10^o$. The solid lines correspond to tailored lithium targets and the dashed lines to foil targets, for various laser powers as noted in the legend.}
\label{fig:fig6}
\end{figure}

\par So far, we have quantitatively described the laser generated $\gamma$-ray flash at various laser powers and for various target materials, demonstrating that $\kappa_\gamma$ can reach significantly high values if the laser background field is not suppressed. Here, the absolute value of $\kappa_\gamma$ is obtained by integrating $\gamma$-photons emitted in a $4 \pi$ solid angle. It has been previously demonstrated for a linearly polarized laser that either by employing specific targets (cylindrical channel \cite{2016_StarkDJ, 2020_WangT} or tightly focused lasers \cite{2022_HadjisolomouP}, $\gamma$-photon emission is stronger in two directions along the polarization plane, forming a double-lobe pattern. We generalize these results, showing that the double-lobe pattern is a general characteristic of linearly polarized laser-matter interactions, as shown in Fig. \ref{fig:fig6}(\textit{A}). The figure shows the $\gamma$-photon radiant intensity, $I_\Omega$ (radiant energy per unit time per solid angle), by approximating that the $\gamma$-photon emission duration equals to the laser pulse duration.

\par The double-lobe feature is observed either for tailored or for foil targets, and at all laser power levels. The main difference among the various cases is the $\gamma$-ray flash amplitude, as seen in Fig. \ref{fig:fig6}(\textit{B}) showing a line-out of the $\gamma$-photon radiant intensity, and by considering photons emitted within $10^o$ full-angle with respect to the polariztion plane. The peak of  Fig. \ref{fig:fig6}(\textit{A}) is slightly higher than the peak of  Fig. \ref{fig:fig6}(\textit{B}) due to the $10^o$ averaging of the radiant intensity.

\par
\begin{figure}[h]
\centering
\includegraphics[width=0.97\linewidth]{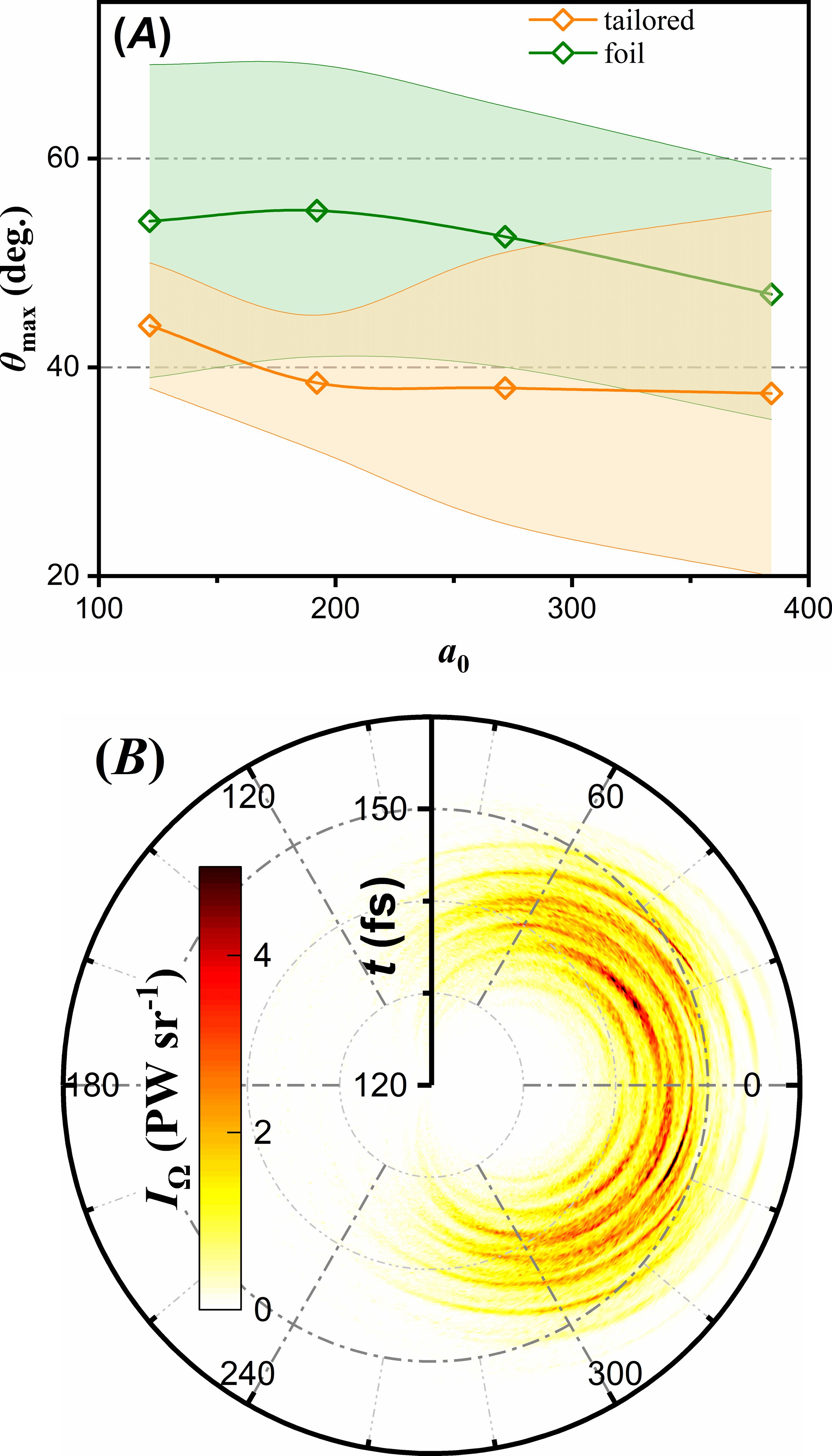}
\caption{(\textit{A}) Angle of the peak $\gamma$-photon radiant intensity as a function of $a_0$. The tailored and foil target cases are shown by the orange and green lines respectively. The ranges for both cases show the region where the radiant intensity is above $90 \kern0.2em \%$ of its peak value. (\textit{B}) Radiant intensity of $\gamma$-photons measured within a full-angle divergence of $10^o$ with respect to the polarization plane, as a function of time-of-flight. In the figure, for computational reasons, only $\gamma$-photons with energy larger than $5 \kern0.2em \mathrm{MeV}$ are considered.}
\label{fig:fig7}
\end{figure}

 Mapping the $\gamma$-photon radiant intensity is crucial in experiments aiming on $\gamma$-photon production through laser-matter interactions. If one needs to optimize $\gamma$-photon detection, the detection system must be aligned along the highest radiant intensity direction, which in all cases lies on the laser polarization plane. However, the polar angle depends on the laser-matter interaction parameters, where the peak can be seen in Fig. \ref{fig:fig7}(\textit{A}) for tailored targets (orange line) and foil targets (green line). The ranges on the plot define where at-least $90 \kern0.2em \%$ of the peak signal is detectable. At all laser power levels, the tailored targets compared to foil targets result in $\gamma$-photon peak emission closer to the laser propagation axis, reaching ${\sim} \kern0.1em 37^o$ versus ${\sim} \kern0.1em 47^o$ respectively for a ${\sim} \kern0.1em 10 \kern0.2em \mathrm{PW}$ laser. Furthermore, in both target cases the peak emission angle moves closer to the laser propagation axis for increased laser power.

\par The $\gamma$-ray flash is usually treated as a time-integrated quantity of a single $\gamma$-photon pulse. However, ballistic evolution of the $\gamma$-photons allows an angular and temporal discrimination, as shown in Fig. \ref{fig:fig7}(\textit{B}); the temporal axis has an arbitrary (detection distance related) offset, not relevant to revealing the $\gamma$-flash dynamics. As seen in the figure, $\gamma$-photon emission is directly connected to the laser wavelength in the regions of strongest emission. Temporal discrimination of the $\gamma$-ray flash reveals that the localized radiant intensity can reach values as high as ${\sim} \kern0.1em 6 \kern0.2em \mathrm{PW \kern0.1em sr^{-1}}$. In addition, the $\gamma$-photons emitted at symmetric azimuthal angles come with a time difference of half laser period.

\par The reduced $\gamma$-photon production on the laser propagation axis indicates that the $\gamma$-photons do not originate from electrons moving antiparallel to the laser propagation direction, where in that case a peak should had been observed rather than a dip. Instead, the $\gamma$-photons are mostly emitted from electrons moving at a non-zero angle along the laser propagation. The lower energy $\gamma$-photons that are emitted as an afterglow at later times have an antiparallel direction with respect to the laser propagation axis and are characterized by a larger period. If a circular detector array is centered at the laser focal spot, then the $\gamma$-ray flashes will be detected with a time delay, as indicated by the non-centered $\gamma$-photon fronts of Fig. \ref{fig:fig7}(\textit{B}), indicating that the major $\gamma$-photon emission happens at a shifted position, at a certain depth inside the target cavity.


\section*{Conclusions}

\par We present the efficient generation of $\gamma$-ray flash through simulations of an ultrashort ultraintense laser interacting with a solid target, where both the background field and main pulse are considered. The effect of the background field is taken from previously published MHD simulation results. The MHD simulations exhibit a unique favorable tailoring of the initially flat target, drilling a conical-like cavity in its volume preceded by an exponentially increasing preplasma distribution. The resulting electron and ion number densities are used as initial conditions for PIC simulations, revealing the effect of the main laser pulse on the tailored target.

\par The conical-like cavity formation strongly alters the laser interaction compared to the interaction with a foil target. The laser pulse, being of similar temporal extent to the cavity depth is reflected by the cavity walls, increasing the laser field intensity by an order of magnitude. Furthermore, cavity reflection of the laser field allows appearance of a longitudinal electric field component, aiding in further propagation of the cavity. In addition, if a relatively low density solid is chosen as the target material, then the current multi-PW lasers are relativistically transparent to the target, penetrating the target volume more deeply.

\par The ultrahigh intensities reached are within the regime where nonlinear Compton scattering process dominates the laser-target interaction, being capable of generating a bright $\gamma$-ray flash. Although no $\mathrm{e^- \mbox{-} e^+}$ pairs were expected from currently available lasers, we estimate that a significant number of pairs can still be produced due to the multiphoton Breit-Wheeler process by the significant intensity enhancement. It is found that the emitted $\gamma$-photons can have an energy approximately equal to the one third of the initial laser energy, with $\gamma$-photon energies approaching the GeV-level, if a ${\sim} \kern0.1em 10 \kern0.2em \mathrm{\mu m}$ thick lithium foil is used as a target and by introducing an appropriate background field.

\par The target material is of crucial importance, since a denser material shifts the interaction out of the relativistic transparency regime. In addition, a denser material results in a shallower cavity, that can not effectively intensify the main laser pulse. As a result, both dense and/or foil targets suppresses the $\gamma$-ray yield since the laser field is strongly reflected, which can also pose a higher risk for optical damage of the laser system.

\par It is found that the $\gamma$-photon radiant intensity forms a double-lobe pattern emitted in the polarization plane, which is a general feature of linearly polarized lasers interacting with matter. The $\gamma$-photon emission peaks at angles of $37^o-55^o$ under the parameters examined. In all cases, the emission angle is smaller when the cavity is present in the target.

\par The temporal dynamics of the $\gamma$-ray flash is also revealed, exhibiting direct connection to the laser period. It is found that for azimuthally symmetrical angles the $\gamma$-photon emission is temporally shifted by half laser period and is suppressed on the laser propagation axis. This pattern indicates that $\gamma$-photon emission originates from electrons co-moving with the laser pulse at a certain angle. Those electrons moving exactly parallel to the pulse produce little-to-no $\gamma$-photons, where this behavior explains the double-lobe pattern observed. In addition, the temporal discrimination of the $\gamma$-ray flash reveals radiant intensities significantly higher than previously expected, opening the road to their application in studying astrophysical processes and to effects of photonuclear interations.





\acknow{The authors would like to thank Ilia P. Tsygvintsev and Vladimir A. Gasilov for providing the MHD simulation results \cite{2022_TsygvintsevIP} and for fruitful discussions. The authors also acknowledge useful communication with Pavel Sasorov on MHD simulations, Martin Matys on computational aspects, Kazuo Tanaka and Domenico Doria on discussing the ELI-NP laser parameters.This work is supported by the projects High Field Initiative (CZ.02.1.01/0.0/0.0/15\_003/0000449) from the European Regional Development Fund. The EPOCH code is in part funded by the UK EPSRC grants EP/G054950/1, EP/G056803/1, EP/G055165/1 and EP/M022463/1.}

\showacknow{}

\bibliographystyle{ieeetr}
\bibliography{pnas-sample}

\end{document}